\documentclass[twocolumn,showpacs,preprintnumbers,amsmath,amssymb]{revtex4}

\usepackage{epsfig}
\usepackage{epsf}
\usepackage{graphics}
\usepackage{graphicx}

\begin{document}

\newcommand{\bc}{\begin{center}}
\newcommand{\ec}{\end{center}}
\newcommand{\be}{\begin{equation}}
\newcommand{\ee}{\end{equation}}
\newcommand{\bq}{\begin{eqnarray}}
\newcommand{\eq}{\end{eqnarray}}
\newcommand{\ikl}{\int_k^\Lambda}
\newcommand{\dq}{\frac{d^4k}{(2 \pi)^4}}
\newcommand{\dn}{\frac{d^nk}{(2 \pi)^n}}
\newcommand{\PLB}{{\it{Phys. Lett. {\bf{B}}}}}
\newcommand{\NPB}{{\it{Nucl. Phys. {\bf{B}}}}}
\newcommand{\PRD}{{\it{Phys. Rev. {\bf{D}}}}}
\newcommand{\AOP}{{\it{Ann. Phys. }}}
\newcommand{\MPL}{{\it{Mod. Phys. Lett. }}}
\newcommand{\del}{\partial}
\newcommand{\g}{\Big( \frac{g}{4 \pi}\Big)}

\title {Consistent momentum space regularization/renormalization of
supersymmetric quantum field theories: the three-loop $\beta$-function for the
Wess-Zumino model}

\date{\today}

\author{David Carneiro}
\email []{david@fisica.ufmg.br}
\author{A. P. Ba\^eta Scarpelli}
\email []{scarp@fisica.ufmg.br}
\author{Marcos Sampaio}
\email []{msampaio@fisica.ufmg.br}
\author{M. C. Nemes}
\email []{carolina@fisica.ufmg.br}

\affiliation{Federal University of Minas Gerais -
Physics Department - ICEx
P.O. BOX 702, 30.161-970, Belo Horizonte MG - Brazil}

\begin{abstract}
\noindent
We compute the three loop $\beta$ function of the Wess-Zumino model  to motivate
implicit regularization (IR) as a consistent and practical momentum-space
framework to study supersymmetric quantum field theories.  In this framework
which works essentially in the physical dimension of the theory we show that
ultraviolet are clearly disantangled from infrared divergences. We obtain
consistent results which motivate the method as a good choice to study
supersymmetry anomalies in quantum field theories.
\end{abstract}
\maketitle

\section{Introduction}

Dimensional regularization (DR) is a remarkable framework which besides
preserving gauge invariance  is relatively simple from the computational
standpoint.  For this reason it has become the stantard method in perturbative
calculations in quantum field theory.   The regularization and renormalization
of supersymmetric gauge theories is, however, a more involved problem.
Momentum space subtraction schemes such as BPHZ \cite{BPHZ} although
supersymmetric are not gauge invariant \cite{PIGUET}. DR does not mantain  the
balance between bosonic and fermionic degrees of freedom due to the analytic
continuation on the space-time dimension and therefore breaks supersymmetry.
Such breaking demands    the calculation of compensating supersymmetry restoring
counterterms. A practical useful modification of DR is dimensional reduction
(DRed) \cite{SIEGEL} which is however mathematically inconsistent and cannot
work at all orders.   As long as such inconsistencies can be tamed  and symmetry
restoring counterterms can be unambiguously generated by imposing the validity
of the Slavnov-Taylor and Ward identities \cite{HOLLIK} the use of DR and DRed
are obviously justified. However it is not always the case that the symmetry of
the Lagrangian is still a symmetry of the full quantum effective action.
Supersymmetry anomalies can in principle  be generated and some erroneous claims
about their existence have occurred because it is difficult to distinguish
between a genuine anomaly and an apparent violation of a supersymmetric Ward
identity due to the use of an ill-defined regularization scheme \cite{JACK}. In
the case of a genuine anomaly no symmetry restoring (finite) counterterms can be
obtained.

A regularization/renormalization framework  that: a) does not modify the field
theoretical content of the bare Lagrangian (and hence does not
unnecessarily complicate the Feynman rules); b) works in the physical dimension
of the model; c) preserves gauge invariance and
d) is friendly from the calculational viewpoint, is therefore desirable to
tackle the issues that we discussed above.    In this sense two
relatively new frameworks deserve special attention: Differential and Implicit
regularization/renormatization (DfR and IR, respectively).

DfR is a coordinate space method that defines the correlation functions without
introducing a regulator or counterterms \cite{D1}-\cite{D11}.  A well defined
prescription that (in a minimal sense)  extends product of distributions into a
distribution automatically delivers renormalized finite amplitudes.  The latter
contains an arbitrary mass scale which must be introduced by dimensional reasons
and plays the role of a renormalization group scale. Gauge invariance may be
sistematicaly implemented in a constrained version of DfR, at least to one loop
order. Finally contact with momentum space is made by means of Fourier
transforms.

On the other hand IR is a momentum space scheme where
scattering amplitudes with fixed external momenta are constructed in the first
place. In section \ref{sec:ir} we briefly
outline this method.

The purpose of this paper is to motivate IR as a sound, symmetry preserving
and practical  framework to study dimension specific
theories, among which chiral, topological and supersymmetric gauge theories are
of prime interest. We use the massless Wess-Zumino model as a testing ground for
the consistency of IR in preserving supersymmetry. We present the
Feynman diagram calculation of this model in sections \ref{sec:1l},
\ref{sec:2l} and \ref{sec:3l}. As a  nontrivial check we compute in section
\ref{sec:bf} the $\beta$-function to three loop order and verify the agreement
with other consistent methods. As a by-product we show how infrared and
ultraviolet divergences are clearly disantangled (in opposite to dimensional
methods).  This is important  in the study of certain supersymmetry anomalies.
We also define what is meant by a minimal subtraction within IR and compare with
DR and DfR. Finally, in section \ref{sec:cr},  we address some theoretical and
phenomenological problems where IR could be useful.

\section{Implicit Regularization/Renormalization}
\label{sec:ir}

The main idea behind IR is to isolate  the divergences from
an amplitude as irreducible loop integrals (ILI) which do not depend upon the
external momenta by judiciously using the identity:
\be
\frac{1}{[(k+k_i)^2 - m^2]} = \sum_{j=0}^{N} \frac{(-1)^j (k_i^2 + 2 k_i
\cdot k)^j}{(k^2-m^2)^{j+1}[(k+k_i)^2-m^2]^j} \, .
\label{eqn:exp}
\ee
In the equation above, $k_i$ are the external momenta and $N$ is chosen so that
the last term is finite under integration over $k$. For instance take the
logarithmically divergent integral in four dimensions
\be
\Gamma (p^2) = \int \frac{d^4 k}{(2 \pi)^4} \, \frac{1}{(k^2-m^2)[(k+p)^2-m^2]}
\, . \ee
We use (\ref{eqn:exp})  with $N=1$ to write
\bq
\Gamma (p^2) &=& \int   \frac{d^4 k}{(2 \pi)^4}  \frac{1}{(k^2-m^2)^2}  -
\nonumber \\ &-&  \int   \frac{d^4 k}{(2 \pi)^4}
\frac{p^2 + 2 p \cdot k}{(k^2-m^2)^2[(k+p)^2-m^2]}  \, .
\eq
Note that the second term on the rhs of the equation above is finite whereas
the first is an  ILI, namely,
\be
I_{log} (m^2) \equiv   \int   \frac{d^4 k}{(2 \pi)^4}  \frac{1}{(k^2-m^2)^2}  \,
,
\label{eqn:ilog}
\ee
which characterizes the ultraviolet behaviour of the amplitude and need
not to be explicitly evaluated. They can be fully absorbed in the
definition of the renormalization constants. In order to define a mass
independent scheme (or in the case of massless theories) we use  the identity
(\ref{eqn:scale}) which we introduce in section \ref{sec:1l}. Such identity
introduces naturally an arbitrary scale which plays the role of renormalization
group scale.  Infinities of higher order are equally displayed as ILI such as
$I_{quad} (m^2)$, $I_{lin} (m^2)$ etc.,  for quadratic and linearly divergent
integrals, respectively. Local arbitrary counterterms can show up in IR
as (finite) differences between divergent integrals of the same superficial
degree of divergence which can sistematically be cast into a set of
``consistency
relations" according to the space-time dimension. At the one loop level they are
related to momentum routing invariance in the loop of a Feynman diagram. A
constrained version of IR corresponds to setting such consistency relations to
vanish. In this way,  abelian and nonabelian gauge invariance can be shown to
automatically implemented.  Ultimately, in a more general context, such local
arbitrary counterterms parametrized by the consistency relations should be left
arbitrary till the end of the calculation when the symmetry content of the
underlying model  may fix its value.    We address the reader to ref.
\cite{IR1}-\cite{IR6} for details and applications of IR.

Overlapping divergences which are the chief complication in any renormalization
scheme can also be handled in a schematic way within IR. This was
illustrated to $n$-loop order  \cite{CAROLINA} in the context of $\phi^3$ theory
in 6 dimensions.

\section{Perturbation expansion of the Wess-Zumino Model}
\label{sec:pt}

We closely follow the notation  and the conventions of ref. \cite{ABBOTT}. The
Wess-Zumino (WZ) model superspace action reads
\bq
{\cal{S}} &=& \int d^4 x \,\, \Bigg\{ \int d^2 \theta d^2 \bar{\theta} \,\,
\bar{\phi}_0 \phi_0 \nonumber \\ &-& \frac{g_0}{3!} \Bigg( \int  d^2 \theta
\phi_0^3 -   \int d^2 \bar{\theta} \bar{\phi}_0^3  \Bigg) \Bigg\} \, ,
\eq
where $(x^a, \theta^\alpha, \bar{\theta}^{\dot{\alpha}})$, $a=1, 2,3,4$,
$\alpha = + -$, $\dot{\alpha}= \dot{+}, \dot{-}$ are coordinates of $d=4$, $N=1$
superspace and $\phi$ ($\bar{\phi}$) is a chiral (antichiral) superfield.

\begin{figure}[h]
  \centerline{
    \epsfxsize=1.3in
    \epsffile{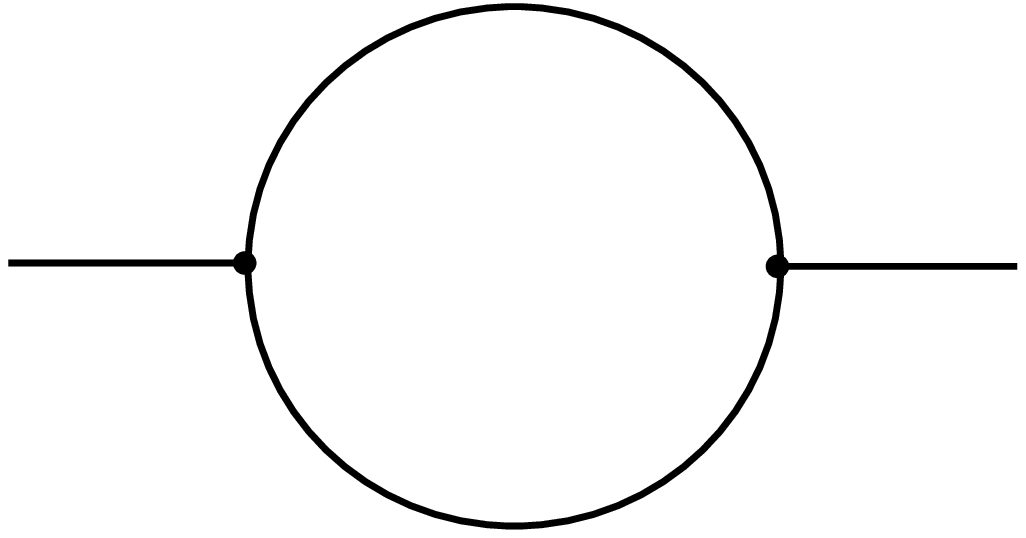}
             }
  \caption[]{\label{fig1} $F(p^2)$}
\end{figure}

\begin{figure}[h]
  \centerline{
    \epsfxsize=1.3in
    \epsffile{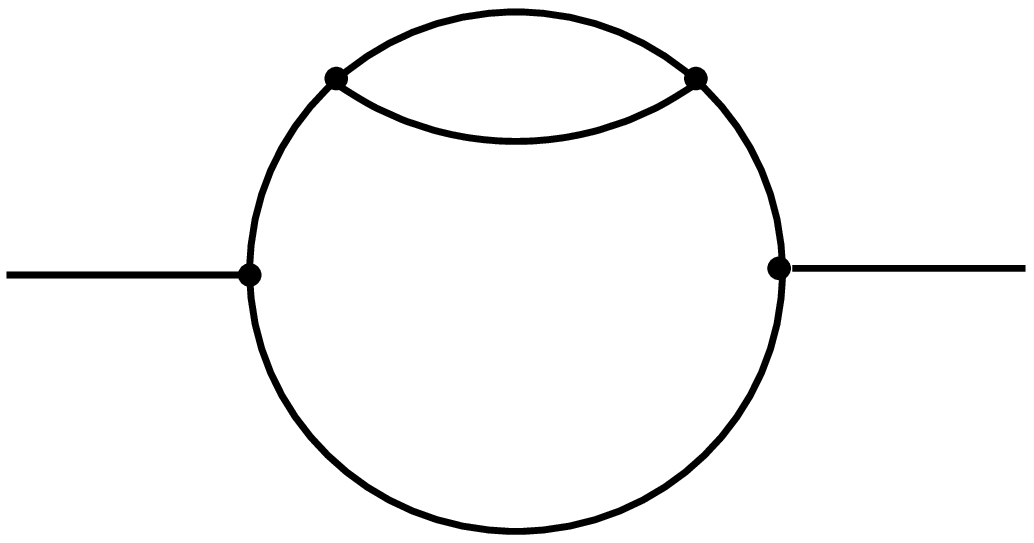}
             }
  \caption[]{\label{fig2} $G(p^2)$}
\end{figure}

\begin{figure}[h]
  \centerline{
    \epsfxsize=1.3in
    \epsffile{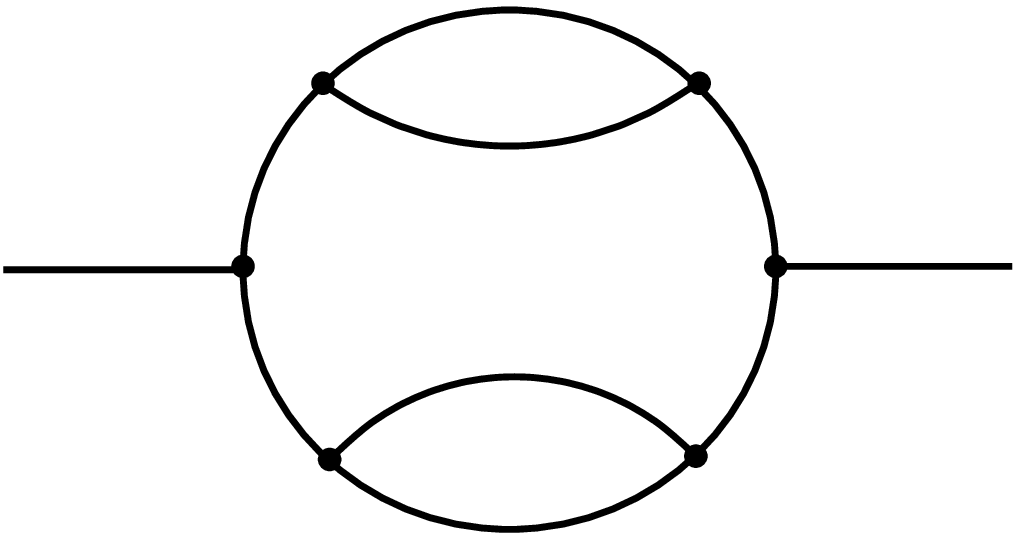}
             }
  \caption[]{\label{fig3} $H_3 (p^2)$}
\end{figure}

\begin{figure}[h]
  \centerline{
    \epsfxsize=1.3in
    \epsffile{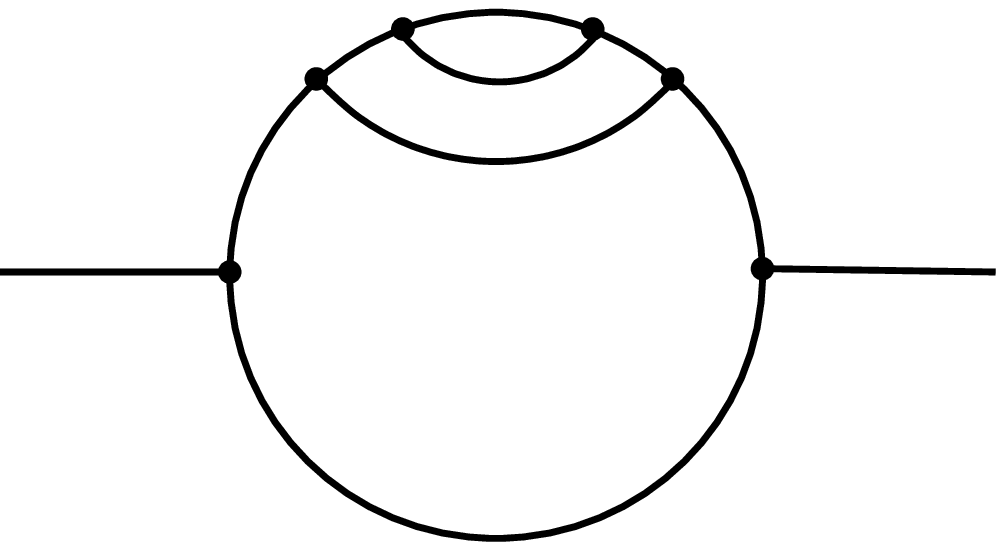}
             }
  \caption[]{\label{fig4} $H_4 (p^2)$}
\end{figure}

\begin{figure}[h]
  \centerline{
    \epsfxsize=1.3in
    \epsffile{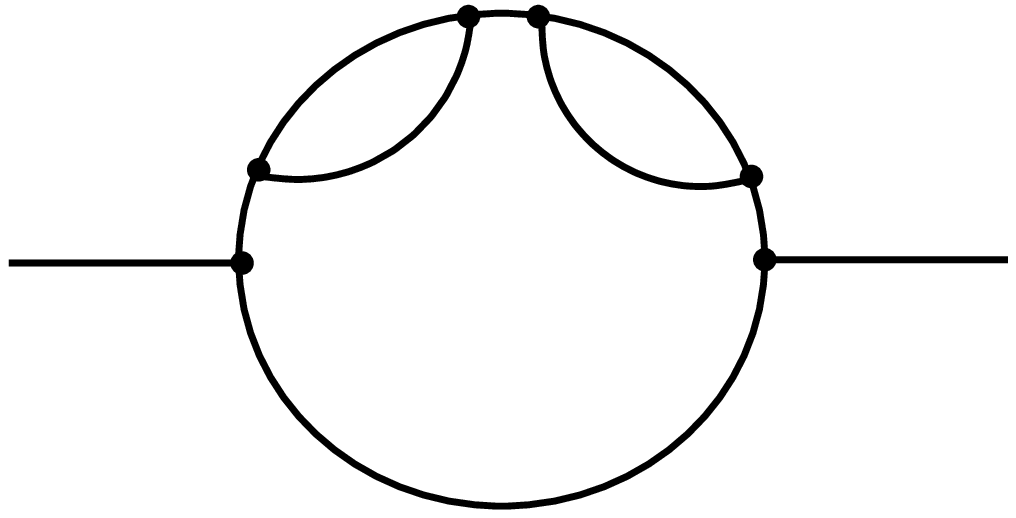}
             }
  \caption[]{\label{fig5} $H_5 (p^2)$}
\end{figure}

\begin{figure}[h]
  \centerline{
    \epsfxsize=1.3in
    \epsffile{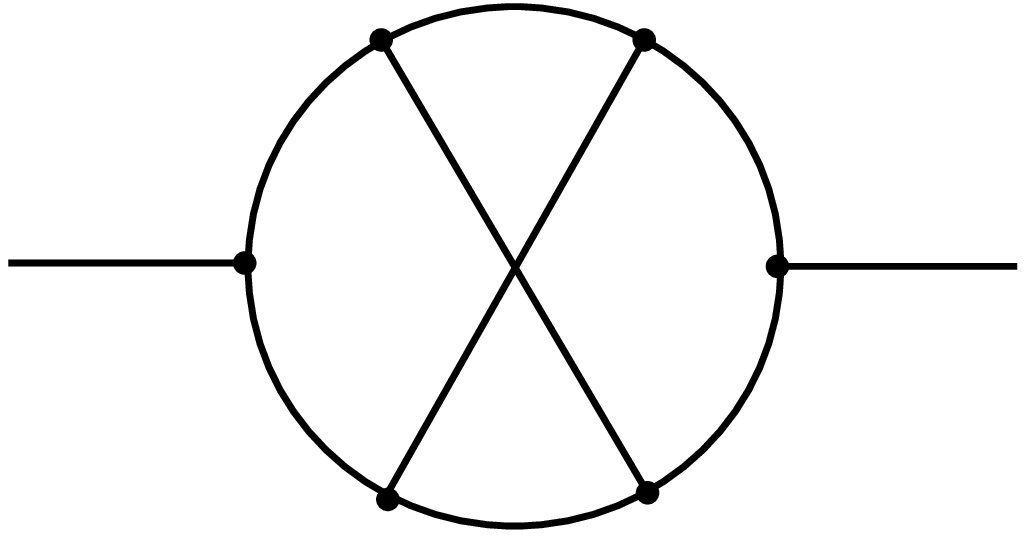}
             }
  \caption[]{\label{fig6} $H_6 (p^2)$}
\end{figure}

The Lagrangian can be written  in terms of component fields
\bq & &\phi_0 = \exp{ \Big( i
\theta^\alpha \bar{\theta}^{\dot{\beta}} \,\, \sigma^a _{\alpha \dot{\beta}}
\,\,  \partial_a \Big) } \times \nonumber \\ & & \times \Big(   1/\sqrt{2}
(A+iB) + \theta^\alpha \psi_\alpha + 1/\sqrt{2} (F-iG) \theta^2
 \Big)\, ,
\eq
which after eliminating the auxiliary fields $F$, $G$, will have the form
\bq
\cal{L} &=& - \frac{1}{2} (\partial_\mu A)^2 -    \frac{1}{2} (\partial_\mu B)^2
- \frac{1}{2} \bar{\psi} \gamma \cdot \partial \psi \nonumber \\
&-& \frac{1}{16} g_0^2 (A^2 + B^2)^2 + \frac{g_0}{2 \sqrt{2}} \bar{\psi}(A + i
\gamma_5 B) \psi \, .
\eq
The WZ model involves only
one renormalization constant since only propagator-type diagrams can diverge. To
see that, one may employ the superfield power counting rules described in
\cite{GRISARU} to conclude that the three-point function is finite. Defining the
renormalization constants as
\bq
g_0 &=& Z_g g \nonumber \\
\phi_0 &=& Z_\phi^{1/2} \phi \, ,
\eq
and using that $g_0 \phi^3_0$ is finite (and hence $Z_g Z_\phi^{3/2} =1$)
enables us to write
\be
g_0 = Z_\phi^{-3/2} g \, .
\label{eqn:g}
\ee
Thus for computing  the $\beta$-function it is sufficient to calculate
$Z_\phi$ (i.e. the divergent structure of the two-point function). We
expand $Z_\phi$    into a power series in the coupling constant,
\be
Z_\phi = 1 + \Big(  \frac{g}{4 \pi} \Big)^2 Z_1 +   \Big(  \frac{g}{4 \pi}
\Big)^4 Z_2+ \Big(  \frac{g}{4 \pi} \Big)^6 Z_3  + \cdots
\label{eqn:Zphi}
\ee
and write the part of the effective action which is linear in $\phi$ and
$\bar{\phi}$ like
\be
\Gamma_2(\phi , \bar{\phi}) = - i \int_{k, \theta,\bar{\theta}}
\bar{\phi}(-k,\theta,\bar{\theta}) \phi (k,\theta,\bar{\theta}) \Delta^{-1} (k)
\, ,
\ee
where $\int_k$ and $\int_\theta$ stand for $\int d^4 k /(2
\pi)^4$ and  $\int d^2 \theta$, respectively. Following \cite{ABBOTT} we notice
that in the renormalized Lagrangian $Z_\phi$ appears only in the kinetic term
${\bar{\phi}}_0 \phi_0 = Z_\phi {\bar{\phi}} \phi$ which amounts to introducing
a factor of $Z_\phi^{-1}$ for each propagator in the diagrams. Hence we can
write
\bq
\Delta^{-1} (p)  &=& Z_\phi + \frac{1}{2} \g^2 Z_\phi^{-2} F (p^2) +
\frac{1}{2} \g^4 Z_\phi^{-5} G (p^2) \nonumber \\
&+&  \g^6 Z_\phi^{-8} H(p^2)  +
\cdots    \, ,
\label{eqn:2PF}
\eq
in which $\frac{1}{2} g^2/(4 \pi)^2 F (p^2) $ represents the contribution of the
one loop diagram (figure \ref{fig1}),
$\frac{1}{2} g^4/(4 \pi)^4 G (p^2) $   represents the two loop
contribution (figure \ref{fig2}),  $\frac{1}{2}$ being a
symmetry factor, and    $g/(4 \pi)^6 H (p^2)$ refers to the
three loop diagrams (figures \ref{fig3} to \ref{fig6}).  Now we take equation
(\ref{eqn:Zphi}) in equation (\ref{eqn:2PF})  and reorganize the power
expansion to write
\bq
&& \Gamma_2(\phi , \bar{\phi}) = - i \int_{p, \theta,\bar{\theta}}
\bar{\phi}(-p,\theta,\bar{\theta}) \phi (p,\theta,\bar{\theta})  \Bigg\{
 Z_\phi + \nonumber \\ && + \frac{1}{2} \g^2 F(p^2)
+ \g^4 \Big[ \frac{1}{2} G(p^2) - Z_1 F(p^2)  \Big]  + \nonumber \\
&& + \g^6 \Big[ H(p^2) - \frac{5}{2} Z_1 G (p^2) -
\Big( Z_2 - \frac{3}{2} Z_1^2 \Big) F(p^2) \Big] + \nonumber \\ && + O\Big( \g^8
\Big) \Bigg\}
\label{eqn:main}
\eq

The Feynman rules for the Wess-Zumino model are well-known (please see
\cite{ABBOTT}, \cite{SRIVASTAVA}) so we shall not derive them here.
The physical interpretation of equation (\ref{eqn:main})  is straightforward. It
defines the renormalization constants $Z_i$'s at each loop order after the
infinities corresponding to subgraphs of previous orders are duly subtracted.  Diagrammatically it
amounts to substituting the divergent subdiagrams of a diagram with their finite
part order by order, as we shall see in the next sections. We shall
perform such substitution after subtracting the divergences in a  minimal sense
within our approach, namely by subtracting irreducible loop integrals.



\section{One loop contribution}
\label{sec:1l}

The one loop contribution to the propagator correction  is represented by the
diagram depicted in figure \ref{fig1}. Application of the Feynman rules to this
diagram  yields \cite{ABBOTT}, \cite{GRISARU}:
\be
\Gamma_2^{(1)} = (-i)^2  g^2  \int_{p} d^4 \theta \,  \bar{\phi}(-p,\theta) \phi
(p,\theta)\int_k \frac{1}{k^2 (p+k)^2}\, .
\ee
From the equation above  and (\ref{eqn:main}) we identify
\be
F(p^2)= - i (4 \pi)^2 \int_k \frac{1}{(k^2-\mu^2) [(p+k)^2-\mu^2]}
\ee
in which we have introduced an infrared cutoff $\mu$. Following our approach, an
irreducible loop integral can be separated from the amplitude above  with
the help of equation (\ref{eqn:exp})  to give
\bq
F(p^2) &=&    - i (4 \pi)^2  \Bigg(   \int_k \frac{1}{(k^2-\mu^2)^2} - \nonumber
\\ &-& \int_k \frac{p^2 + 2 p \cdot k}{(k^2 -\mu^2)^2[(p+k)^2 - \mu^2]} \Bigg)
\nonumber \\
\eq
$$
\equiv - i (4 \pi)^2  \Bigg( I_{log} (\mu^2)  - b \int_0^1 dz \,
\ln \Big( \frac{p^2 z ( z-1)}{ \mu^2} + 1 \Big) \Bigg)\, ,
$$
where, henceforth we define
\be
b \equiv \frac{i}{(4 \pi)^2}
\ee
and $I_{log} (\mu^2)$   is given by equation (\ref{eqn:ilog}). Since  the limit
where $\mu \rightarrow 0$ is infrared ill-defined in $I_{log} (\mu^2)$, the
correct ultraviolet behaviour is obtained by exchanging the latter against
$I_{log} (\lambda^2)$ provided  $\lambda^2 \ne 0$ by means of the identity
\be
I_{log} (\mu^2) =  I_{log} (\lambda^2) + b \ln \Big( \frac{\lambda^2}{\mu^2}
\Big) \, .
\label{eqn:scale}
\ee
Hence  (\ref{eqn:scale})  splits the ultraviolet and
infrared divergences and as a byproduct  it parametrizes the arbitrariness in
separating the divergent from the finite part. That is because the infrared
divergent piece in the right hand side  of      (\ref{eqn:scale})  cancels out
against another piece coming from the (ultraviolet) finite part of the amplitude
which may be easily integrated to give
\be
F(p^2) = - i (4 \pi)^2  \Bigg( I_{log} (\lambda^2)  +b \ln
\Big( \frac{-\lambda^2 e^2 }{p^2} \Big) \Bigg)\, .
\label{eqn:F}
\ee
Consequently $\lambda$ is the natural candidate for a renormalization scale from
which we may construct a Callan-Symanzik renormalization group equation.
A minimal subtraction within our approach amounts to defining, with the help of
(\ref{eqn:main}),
\be
Z_1 = \frac{i}{2}(4 \pi)^2    I_{log} (\lambda^2)  , \quad \lambda^2 \ne 0,
\label{eqn:Z1}
\ee
i.e.,  we have subtracted only the irreducible loop integral.  For future
reference, we define the finite part of $F(p^2)$ as
\be
F_{\mbox{\scriptsize{fin}}} =   - i (4 \pi)^2  b \ln \Big( \frac{-\lambda^2 e^2
}{p^2} \Big)\,
\label{eqn:Ffin}
\ee
where $\lambda$ plays the role of an arbitrary local counterterm. Moreover
$F_{\mbox{\scriptsize{fin}}}$ satisfies a Callan-Symanzik renormalization group
equation with scale $\lambda$ \cite{PRD}.

For a massive theory  a minimal mass independent scheme is defined in a similar
fashion. In  \cite{PRD} we have compared our minimal subtraction scheme with
the MS scheme in dimensional renormalization as well as  in differential
renormalization. The arbitrary scales appearing in each
framework  are related to  each other as we shall discuss in
section \ref{sec:bf} (see also \cite{DUNNE}).

\section{Two Loop Contribution}
\label{sec:2l}

The propagator receives just one two-loop contribution (figure  \ref{fig2})
because the only propagator is the one from $\phi$ to $\bar{\phi}$ (there is no
$\phi \phi$ or $\bar{\phi} \bar{\phi}$ propagators). The corresponding amplitude
can be constructed by direct application of the  Feynman rules. After some
straightforward algebra it can be reduced to
\bq
\Gamma_2^{(2)}  &=& (-i)^3 g^4 \int_p d^4 \theta  \,  \bar{\phi}(-p,\theta) \phi
(p,\theta) \times \nonumber \\ &\times&  (-1) \int_{q,k}  \frac{1}{k^2 q^2 ( k +
q)^2 (p + q)^2} \, ,
\eq
which enables us to identify
\be
G (p^2) =  (4 \pi)^4 \int _{q} \frac{1}{q^2  (p + q)^2} \int_k  \frac{1}{k^2 (k
+ q)^2} \, .
\label{eqn:G}
\ee
Let us have a closer look at the $O(g^4)$ term in equation (\ref{eqn:main})
viz.,
\be
\Big( \frac{g}{4 \pi} \Big)^4 \Big[ Z_2 + \frac{1}{2} G(p^2) - Z_1 F(p^2) \Big]
\, ,
\label{eqn:sub2}
\ee
which graphically corresponds to figure \ref{fig2}. Perturbative
renormalization is inductive in the definition of the counterterms. Thus in
defining $Z_2$ we firstly ought to subtract the subdivergence through  $Z_1$
which was defined in the previous order as prescribed in (\ref{eqn:sub2}). We
can explicitly evaluate  $1/2 \, G(p^2) - Z_1 F(p^2)  $ using  (\ref{eqn:F}),
(\ref{eqn:Z1}) and (\ref{eqn:G}) to conclude that such operation amounts to
substituting the subintegration over $k$  in (\ref{eqn:G})  with
$F_{\mbox{\scriptsize{fin}}}(p^2)$ divided by $- i (4
\pi)^2 $ to define:
\bq
\widetilde{G} (p^2) &=&  b (4 \pi)^4 \int_{q} \frac{1}{q^2 (p + q)^2} \ln \Big(
\frac{- \lambda^2 e^2}{ q^2} \Big) \nonumber \\ &=&
\frac{1}{2} G(p^2) - Z_1 F(p^2) \, .
\label{eqn:GG}
\eq
Hereforth the tilde means that we have subtracted all the subdivergencies.
The graphical interpretation of such procedure is clear.

We proceed to define the renormalization constant $Z_2$. Introducing an infrared
cutoff in (\ref{eqn:GG}) enables us to write
\be
\widetilde{G} (p^2) = b (4 \pi)^4 \int_{q} \frac{1}{(q^2-\mu^2) [(p +
q)^2-\mu^2]} \ln \Big( \frac{- \lambda^2 e^2}{ q^2 - \mu^2} \Big) \, .
\label{eqn:GG1}
\ee
The irreducible loop integral can be separated just as we did at the one loop
level as follows
\bq
\widetilde{G} (p^2) &=& b (4 \pi)^4 \int_{q} \frac{1}{(q^2-\mu^2)^2 }\ln \Big(
\frac{- \lambda^2 e^2}{ q^2 - \mu^2 } \Big)  \nonumber \\
&-& b (4 \pi)^4 \int_{q} \frac{p^2 + 2 p \cdot q}{(q^2-\mu^2)^2 [(p +
q)^2-\mu^2]} \times  \nonumber \\ &\times& \ln \Big( \frac{- \lambda^2 e^2}{
q^2 - \mu^2} \Big)
\label{eqn:GG2}
\eq
Notice that the second term on the right hand side is finite whilst the first
term is an ILI, namely
\be
 \int_{q} \frac{1}{(q^2-\mu^2)^2} \ln \Big(
\frac{- \lambda^2 e^2}{ q^2 - \mu^2} \Big) \equiv   I_{log}^{(2)} (\mu^2)  \, .
\label{eqn:ilog2}
\ee
Again,  a bona fide renormalization
constant should be infrared finite. Thus we subtract $I_{log}^{(2)}
(\lambda^2)$ with $\lambda \ne 0$  by parametrizing the infrared divergence
by means of the identity
\bq
I_{log}^{(2)} (\mu^2) =    I_{log}^{(2)} (\lambda^2)  + b \Bigg[ \frac{1}{2}
\ln^2 \Big( \frac{\mu^2}{\lambda^2 e^2}\Big) - 2 \Bigg]
\label{eqn:scale2}
\eq
Relation (\ref{eqn:scale2}) is the two loop analog of (\ref{eqn:scale}) .
Now taking  (\ref{eqn:scale2}) and    (\ref{eqn:ilog2}) into (\ref{eqn:GG2})
permits us to subtract the divergence by defining the renormalization constant
of order $g^4$   as
\be
Z_2 = - \frac{i}{2} (4 \pi)^2     I_{log}^{(2)} (\lambda^2) \, .
\label{eqn:Z2}
\ee

In the next section we shall evaluate the three-loop contributions to the
propagator and define the corresponding counterterms.   Thus it is convenient to
simplify the finite part of the two loop contribution as there will be two loop
subgraphs at three-loop order.  We show in appendix A that
\bq
G_{\mbox{{\scriptsize{fin}}}}(p^2) &=&
\widetilde{G}_{\mbox{{\scriptsize{fin}}}}(p^2) \nonumber \\ &=&     -
\frac{1}{2} \ln^2 \Big( \frac{- \lambda^2 e^2}{p^2} \Big) - \ln \Big( \frac{-
\lambda^2 e^2}{p^2} \Big)    \, .
\label{eqn:fin2}
\eq
It is worthwhile mentioning  that   $G_{\scriptsize{{\mbox{fin}}}} (p^2)$  is
infrared safe as it should, since the limit where $\mu \rightarrow 0$ is well
defined through a cancellation of terms in  equations (\ref{eqn:GG2}) and
(\ref{eqn:scale2}) .

\section{Three Loop Contributions}
\label{sec:3l}

The diagrams depicted in figures 3 to 6 represent the three loop order
contributions to the propagator. Feynman rules can be directly applied to
give  \cite{ABBOTT} :
\bq
& &\Gamma_2^{(3)} = (-i)^4 g^6 \int_p d^4 \theta \,\,   \bar{\phi}(-p,\theta)
\phi (p,\theta)  \times \nonumber \\ && \frac{-i}{(4 \pi)^6} \Big(
H_3 (p^2) + H_4 (p^2) + H_5 (p^2) + H_6 (p^2) \Big) \, ,
\label{eqn:3loop}
\eq
$H_3 \ldots H_6$  being the result of
integrating over the $\delta$ functions of the superspace coordinates and
eliminating the covariant derivatives. The factor $-i/(4 \pi)^6$ appears so that
the $H_i$'s here agree with the definition expressed in equation
(\ref{eqn:main}). In the latter $H$ stands for $H_3 + \ldots + H_6$.

Take the first three loop diagram (figure \ref{fig3}). It contributes to
(\ref{eqn:3loop}) with
\bq
H_3 &=& \frac{i (4 \pi)^6}{8} \int_q \frac{1}{q^2 (p+q)^2}\int_k \frac{1}{k^2
(k+p+q)^2} \times \nonumber \\ &\times& \int_l \frac{1}{l^2 (q+l)^2} \, .
\label{eqn:H3}
\eq
Note that it contains two one-loop  subdiagrams. According to what we discussed
earlier we can substitute the integral over $l$ with
$F_{\scriptsize{\mbox{fin}}}(q^2)/[-i (4 \pi)^2]$ (equation (\ref{eqn:Ffin}))
at once.  As for the integral over $k$ , it can be cast with the help of
(\ref{eqn:exp}) as
\bq
&&\frac{i (4 \pi)^6}{8} \Bigg[ \int_q \frac{1}{q^2 (p+q)^2}\Bigg( \int_k
\frac{1}{k^2 (k+q)^2} - \nonumber \\ &-&
\frac{p^2 + 2 p \cdot (k+q) }{k^2 (k+q)^2 (p+k+q)^2} \Bigg) b \ln \Big(
\frac{-\lambda^2 e^2}{q^2} \Big) \Bigg]       \, .
\label{eqn:H31}
\eq
This is useful because doing so we have contributed to free the ultraviolet
divergent piece from external momentum  dependence. Note that only the first
piece of equation (\ref{eqn:H31}),  namely
\be
\frac{i (4 \pi)^6}{8} \int_q \frac{1}{q^2 (p+q)^2} \int_k
\frac{1}{k^2 (k+q)^2}  b \ln \Big(
\frac{-\lambda^2 e^2}{q^2} \Big) \,
\label{eqn:H32}
\ee
is  ultraviolet divergent. Thus we may promptly substitute the integral over $k$
(which represents a one-loop subdiagram with external
momentum $q$)  in (\ref{eqn:H32}) with  $F_{\scriptsize{\mbox{fin}}}(q^2)/[-i (4
\pi)^2]$ to define:
\be
\widetilde{H}_3 (p^2) = \frac{i (4 \pi)^6}{8} \int_q \frac{1}{q^2 (p+q)^2}
\Bigg[ b \ln \Big( \frac{-\lambda^2 e^2}{q^2} \Big) \Bigg]^2\, .
\ee
To define the irreducible loop integral which will
contribute to $Z_3$  we introduce an infrared cutoff $\mu$ and remove the
external momentum dependence in $\widetilde{H}_3 (p^2)$ in a similar fashion to
equations (\ref{eqn:GG1})-(\ref{eqn:GG2}). That is to say
\bq
\widetilde{H}_3 (p^2) &=& \frac{i (4 \pi)^6}{8} \int_q \frac{1}{(q^2-\mu^2)^2}
b^2 \ln^2 \Big( \frac{-\lambda^2 e^2}{q^2-\mu^2} \Big) + \nonumber \\
&+& {\cal{F}}_3 (p^2,\mu^2)  \nonumber \\ &=&
- \frac{i (4 \pi)^2}{8} I_{log}^{(3)} (\mu^2)  +  {\cal{F}}_3 (p^2,\mu^2) \, ,
\eq
in which we defined another logarithmically divergent ILI,
\be
I_{log}^{(3)} (\mu^2)  \equiv \int_q \frac{1}{(q^2 - \mu^2)^2} \ln^2
\Big( \frac{-\lambda^2 e^2}{q^2 - \mu^2} \Big) \, ,
\label{eqn:ilog3}
\ee
whereas   $ {\cal{F}}_3 (p^2,\mu^2) $ stands for an ultraviolet finite piece.
This diagram will contribute to $Z_3$ with an infrared finite term
\be
Z_3^{(3)} = \frac{i (4 \pi)^2}{8}  I_{log}^{(3)} (\lambda^2) \, ,
\ee
$\lambda \ne 0$ since one can easily verify that
\be
I_{log}^{(3)} (\mu^2) = I_{log}^{(3)} (\lambda^2) - b \Big[
\frac{8}{3} + \frac{1}{3} \ln^3 \Big( \frac{\mu^2}{\lambda^2 e^2} \Big) \Big] \,
.
\label{eqn:scale3}
\ee
In analogy with our calculations at the one and two loop orders,  the
infrared divergent piece in equation above is expected to cancel  the
infrared divergence in ${\cal{F}}_3 (p^2,\mu^2)$ to render a  well
defined finite part. This is indeed the case as one can prove
after some straightforward algebra. Therefore we may write
\be
\widetilde{H}_3 (p^2)  = - \frac{i (4 \pi)^2}{8} I_{log}^{(3)} (\lambda^2)  +
{\cal{G}}_3 (p^2,\lambda^2) \, .
\ee

The second three-loop contribution is represented by figure \ref{fig4}
from which one gets
\bq
H_4 &=&   \frac{i (4 \pi^6)}{2} \int_{q} \frac{1}{q^2 (p+q)^2}
\int_ l \frac{1}{l^2 (q+l)^2} \times \nonumber \\ &\times&
\int_k \frac{1}{k^2 (k+l)^2}     \, .
\eq
One can see from this graph that it contains as a subdiagram
the two loop graph shown in figure \ref{fig2} which is represented by the
integrals over $l$ and $k$ in the amplitude above, $q$ playing the role of
external momentum.  The procedure is identical as  we did for $H_3$ so we
shall only summarize the steps.  We replace the integrals over $l$ and $k$ by
$G_{\mbox{{\scriptsize{fin}}}}(q^2)/(4 \pi)^4$ (equation (\ref{eqn:fin2})) and
introduce an infrared cutoff, $\mu$. Next we expand the propagator which
contains the external momentum $p$ in the usual fashion so to define an
irreducible loop integral which is $p$-independent. We obtain
\bq
\widetilde{H}_4 (p^2) &=& \frac{-i (4 \pi)^6}{2 (4 \pi)^4} \int_q
\frac{1}{(q^2 - \mu^2)^2} \times \nonumber\\
&\times& \Bigg[    \frac{1}{2} \ln^2 \Big( \frac{- \lambda^2 e^2}{q^2-\mu^2}
\Big) + \ln \Big( \frac{- \lambda^2 e^2}{q^2-\mu^2} \Big)    \Bigg] \nonumber \\
&+&  {\cal{F}}_4 (p^2,\mu^2)    \, ,
\eq
${\cal{F}}_4$ being an ultraviolet finite term.  Using (\ref{eqn:ilog2}),
(\ref{eqn:scale2}), (\ref{eqn:ilog3}) and  (\ref{eqn:scale3}) allows us to write
\bq
\widetilde{H}_4 (p^2) &=& \frac{-i (4 \pi)^2}{2} \Big( \frac{1}{2} I_{log}^{(3)}
(\lambda^2)  +  I_{log}^{(2)} (\lambda^2)      \Big)  \nonumber \\ &+&
{\cal{G}}_4 (p^2,\lambda^2) \, ,
\eq
where ${\cal{G}}_4 (p^2,\lambda^2)$, $\lambda^2 \ne 0$,  is now both ultraviolet
and infrared finite . Finally this diagram contributes to
$Z_3$ with
\be
Z_3^{(4)} = \frac{i (4 \pi)^2}{2} \Big( \frac{1}{2} I_{log}^{(3)}
(\lambda^2)  +  I_{log}^{(2)} (\lambda^2)      \Big) \, .
\ee

The diagram displayed in figure \ref{fig5} is easy to evaluate. It also contains
two one-loop subdiagrams  just as the diagram in figure \ref{fig3}. Therefore we
expect that it shall contribute to $Z_3$ with a term proportional to
$I_{log}^{(3)} (\lambda^2)$ as well.  For the sake of completeness we write
its contribution to (\ref{eqn:3loop}):
\bq
H_5 &=&   \frac{i (4 \pi)^6}{4} \int_{q} \frac{1}{q^2 (p+q)^2}
\int_ l \frac{1}{l^2 (q+l)^2} \times \nonumber \\ &\times&
\int_k \frac{1}{k^2 (k+q)^2}     \, .
\eq
Notice that the integrals over $l$ and $k$ in the equation above represent the
one loop subdiagrams. We proceed in a similar fashion as  we did for the diagram
in figure \ref{fig3} to obtain
\be
\widetilde{H}_5 (p^2) = \frac{-i (4 \pi)^2}{4}  I_{log}^{(3)}
(\lambda^2)  + {\cal{G}}_5 (p^2,\lambda^2) \, ,
\ee
where the notation is now obvious. Hence, the renormalization constant
$Z_3$ should contain the following contribution,
\be
Z_3^{(5)} = \frac{i (4 \pi)^2}{4} I_{log}^{(3)} (\lambda^2) \, ,
\ee
in order to cancel the divergence that stemmed from this diagram.

Finally we study the diagram in figure \ref{fig6}. In opposite to the previous
three loop diagrams, it contains no subdiagrams at all. Its contribution can be
cast as    \cite{ABBOTT}:
\be
H_6 =   \frac{i (4 \pi)^6}{2} \int_{l,q,k} \frac{1}{q^2 k^2 ( p - l )^2 (
l - q )^2 ( l - k )^2 (q - k)^2}  \, .
\ee
An irreducible loop integral can be displayed after
expanding the propagator that contains $p$ in $H_6$ according to (\ref{eqn:exp})
to obtain
\bq
H_6 &=&   \frac{i (4 \pi)^6}{2} \int_{l,q,k} \frac{1}{q^2 k^2  l ^2 (
l - q )^2 ( l - k )^2 (q - k)^2} \nonumber \\ &+& {\cal{F}}_6 (p^2) \equiv
H_6^{\infty} + {\cal{F}}_6 (p^2)\, ,
\label{eqn:h6}
\eq
${\cal{F}}_6 (p^2)$ being a ultraviolet finite piece as usual. We show in
appendix B that
\be
H_6 =  - 3 i  (4 \pi)^6 \zeta (3) I_{log} (\lambda^2)   + {\cal{G}}_6 (p^2,
\lambda^2) \, ,
\ee
where $\zeta (x)$ stands for the Riemann zeta-function. Therefore we define the
last contribution to $Z_3$,
\be
Z_3^{(6)} = 3 i  (4 \pi)^6 \zeta (3) I_{log} (\lambda^2) \, ,
\label{eqn:z36}
\ee
to finally write
\be
Z_3 = Z_3^{(3)} + \ldots + Z_3^{(6)}.
\label{eqn:Z3}
\ee

\section{The $\beta$-function}
\label{sec:bf}

In \cite{PRD} we compared implicit, dimensional, differential
and BPHZ renormalization and shown, within these intrinsically distinct
frameworks, how renormalizations schemes and scales are related.  It
goes as follows. Minimally subtracting the infinities in dimensional
regularization (i.e. removing only the poles) delivers a finite
(non-counterterm) term which depends upon the renormalization scale $\mu$. It
results that such term is the same that appears should we employ
differential renormalization, except for a simple rescaling of its typical
arbitrary scale $M^2$ (that is to say, a finite counterterm).  The latter is the
scale of  a Callan-Symanzik renormalization group equation satisfied by the
vertex function \cite{D2}. A minimal renormalization scheme within implicit
regularization corresponds to subtract the irreducible loop integrals in a mass
independent fashion through relations like (\ref{eqn:scale}),
(\ref{eqn:scale2}) and (\ref{eqn:scale3}) where an arbitrary mass scale
$\lambda$ is introduced. It turns out that the resulting finite part can be
identified with the one from differential renormalization (Fourier
transformed to momentum space) after a simple rescaling  of $\lambda$. The
difference is that in our framework we do have counterterms from which we may
calculate the renormalization group functions. Moreover our renormalized
amplitude also satisfies a Callan-Symanzik renormalization group equation
governed by the arbitrary scale $\lambda$.

An important non-trivial test of our framework is to calculate the
$\beta$-function of the WZ model which is well known to four loop
order. We start with equation (\ref{eqn:g}). Because $g_0$ cannot
depend upon our arbitrary scale $\lambda$ we are led to
\be
\beta = \frac{3g}{2}  \lambda \frac{\partial Z_\phi}{\partial \lambda} \equiv
\frac{3g}{2} \dot{Z_\phi} \, ,
\ee
where  the dot stands for the logarithmic derivative, i.e. $\dot{A}
\equiv \lambda (\partial/ \partial \lambda) A$.   Now we revoke (\ref{eqn:Zphi})
to write a coupling constant expansion for the $\beta$ function, viz.
\bq
\frac{\beta}{6 \pi} &=& \dot{Z}_1 \Big(\frac{g}{4 \pi}\Big)^3   +   (\dot{Z}_2
+ 2 Z_1 \dot{Z}_1) \Big(\frac{g}{4 \pi}\Big)^5 + \nonumber \\
&+& (\dot{Z}_3 + 5 \dot{Z}_1 Z_2 + 2 Z_1 \dot{Z}_2 + 4 \dot{Z}_1 Z_1^2 )
\Big(\frac{g}{4 \pi}\Big)^7 +  \nonumber \\
&+& \cdots  \, .
\eq
Using our results displayed in equations (\ref{eqn:Z1}), (\ref{eqn:Z2}) and
(\ref{eqn:Z3})  and that
\bq
\dot{I}_{log} (\lambda^2) &=& - 2 b  \,\, , \nonumber \\
\dot{I}_{log}^{(2)} (\lambda^2) &=& 2  I_{log} (\lambda^2) - 2 b  \,\,\,\,
{\mbox{and}}    \nonumber \\
\dot{I}_{log}^{(3)} (\lambda^2) &=& 4  I_{log}^{(2)} (\lambda^2) - 4 b \, ,
\eq
we get
\bq
\beta &=& g \Bigg[  \frac{3}{2}\Big(\frac{g}{4 \pi}\Big)^2 -
\frac{3}{2}\Big(\frac{g}{4 \pi}\Big)^4  + \nonumber \\
&+&   \Bigg( \frac{21}{4} + 9 \zeta (3) \Bigg)   \Big(\frac{g}{4 \pi}\Big)^6
\Bigg] \, .
\eq
Some comments are in order. Whilst the one and two-loop coefficients of the
$\beta$ function are scheme-independent as opposed to the three loop
coefficient, the coefficient of $\zeta (3)$ should be universal. That is because
the diagram from which such coefficient stems is primitive. Our results for
the $\beta$-function exactly agree with the calculation performed within
differential renormalization \cite{HAAGENSEN}. In contrast, the coefficient we
obtain for $\zeta (3)$ is different from the one obtained in dimensional
regularization in \cite{ABBOTT}. Later, an extension of the work of ref.
\cite{ABBOTT} to four loop order in dimensional regularization \cite{SEN}
calculated the same coefficient of $\zeta (3)$ as the one obtained by us. They
have also verified that their result was in agreement with a consistency
condition which relates the coeficients of $Z_4$ and $Z_3$.  Finally the results
obtained by us here and in differential renormalization \cite{HAAGENSEN} for the
three loop (scheme-dependent) coefficient corresponds to a momentum
space subtraction scheme  (MOM) in which a subtraction is perfomed at
$p^2 = \lambda^2 \ne 0$ \cite{ABBOTT}.

\section{Conclusions and Perspectives}
\label{sec:cr}
If on one hand at the near future  experiments at the LHC
or at a linear $e^+ e^-$ collider  will test decisively supersymmetric
extensions of the standard model, on the other hand  the theoretical machinery
which exploits such predictions must have a thorough control upon the
regularization and renormalization of supersymmetric Yang-Mills theories.

Since complete regularization framework which is both gauge and supersymmetric
invariant has yet not been constructed, it is reasonable to exploit a
framework which is gauge invariant and works in the physical dimension of the
theory. This work is a fundamental step in this direction which should help to
address important issues such as:
\begin{enumerate}
\item gauge field theories with soft
supersymmetry-breaking terms:  for instance QCD with soft breaking which has
a particular   phenomenological interest. In \cite{PIGUET2}  models with
soft-breaking terms have been studied in the Wess-Zumino gauge. However
there appears new parameters which have no clear interpretation as being
either susy or soft breaking terms (see however \cite{HOLLIK}).  IR should be
an useful tool to renormalize softly broken susy gauge theories  avoiding some
complications used  in dimensional methods such as the introduction of
Slavnov-Taylor identities as constraint equations in order to ensure both gauge
invariance and supersymmetry \cite{WIP}.

\item the anomaly puzzle: the axial and energy momentum trace anomalies
seem to lead to the conclusion that  the $\beta$ function of supersymmetric
gauge field theories should be exhausted by the first loop correction
\cite{NSVZ}. Whilst this appear to be the case in models with $N=2$
supersymmetry, the case $N=1$ is less clear \cite{ANOM} particularly when one
uses dimensional regularization methods. In DfR \cite{ANOMDF}, which like IR
does not resort to analytic continuation on the space time dimension, the
infrared origin of the two loop coefficient of the $\beta$-function of $N=1$ SYM
has been discussed (see also \cite{SOLOSHENKO}). IR,  which operates in
momentum space and on the physical dimension of the theory,  clearly isolates
infrared and ultraviolet divergences  by means of distinct scales. Thus we
expect that IR can shed some light on the infrared effects and scheme dependence
of the higher order corrections (if any) to the $\beta$-function of $N=1$
supersymmetric gauge theories as well as establish a correspondence with other
frameworks  such as the canonical Wilsonian coupling flow \cite{MURAYAMA}.
\end{enumerate}


\section*{Appendix A}

We can easily obtain the finite part of $G(p^2)$
in a simple form using Rosner's technique of expanding propagators in
Chebyshev's polynomials \cite{ROSNER}.

In Euclidean space   (\ref{eqn:GG}) reads:
\be
\widetilde{G}_{{\scriptsize{E}}}(p^2)= - (4 \pi)^2 \int_{q_{E}}
\frac{1}{q^2 (p + q)^2} \ln \Big( \frac{\lambda^2 e^2}{ q^2} \Big) \, .
\label{eqn:a1}
\ee
Because  $\widetilde{G}_{E}(p^2)$ is a function of $p^2$ only we may average
over the directions of $p$ in four dimensions.  Let $C_m$ be the  Chebyshev's
polynomial  of order $m$.  It satisfies a four dimensional version
of its orthogonality relation, viz.
\be
\int \frac{d \Omega_p}{2 \pi^2}    C_n
\Big( \frac{p \cdot q}{p q}\Big) C_m
\Big( \frac{p \cdot q}{p q}\Big)  = \delta_{mn} \, ,
\label{eqn:a2}
\ee
with $C_0 = 1$. Following  \cite{ROSNER} we write
\be
\frac{1}{(p+q)^2} = \frac{1}{p q} \sum_{n} \langle p | q \rangle^{n+1} C_n
\Big( \frac{p \cdot q}{p q}\Big) \, ,
\label{eqn:a3}
\ee
where
\bq
\langle p | q \rangle = \left\{ \begin{array}{ll}
                                           p/q & \mbox{if $p < q$}\\
                                           q/p & \mbox{if $p > q$}
                                              \end{array} \right.  \, .
\eq
Using (\ref{eqn:a3}) and (\ref{eqn:a2}) in (\ref{eqn:a1}) yields
\be
\widetilde{G}_{{\scriptsize{E}}}(p^2) = - (4 \pi)^2 \int_{q_{E}} \frac{1}{q^3 p}
\langle q | p \rangle  \ln \Big( \frac{\lambda^2 e^2}{ q^2} \Big) \, ,
\ee
$\int_{q_{E}}  \equiv 1/(2 \pi)^4 \int d \Omega_q \int q^3 \, d q$.
After integrating over the angles we get
\be
\widetilde{G}_{E}(p^2) = - \ln \Big( \frac{\lambda^2 e^2}{p^2} \Big) - 1 - \int_{p^2}^{\infty}
\frac{1}{x} \ln    \Big( \frac{\lambda^2 e^2}{x} \Big) \,\, dx    \, .
\ee
Take the last term in the equation above. That integral contains the
ultraviolet divergent piece from which we define the renormalization constant
$Z_2$. Moreover as we discussed earlier it should be infrared safe. To see that
explicitly  let us split the integral from $p^2$ to $\infty$ in three pieces:
from $p^2$ to $\lambda^2 \ne 0$ plus from $\lambda^2$ to $\mu^2$ (infrared
cutoff) plus a remaining contribution from $\mu^2$ to $\infty$.  The latter
can be easily shown to be related to  (\ref{eqn:ilog2}),
\be
\int_{\mu^2}^{\infty} \frac{1}{x} \ln \Big(
\frac{\lambda^2 e^2}{x} \Big) \,\, dx  = \frac{1}{b}    I_{log}^{(2)} (\mu^2) +
1 \, ,
\ee
whereas the two other pieces added together yield $1/2 \{ \ln^2 [p^2/(\lambda^2
e^2)] -   \ln^2 [\mu^2/(\lambda^2 e^2) ] \}$ . Note that such contribution
diverges as $\mu \rightarrow 0$. However this behaviour is tamed if we
correctly define a genuine ultraviolet divergent object by exchanging $\mu$
against $\lambda \ne 0$ with the help of equation (\ref{eqn:scale2}). Putting
all the results together we arrive at
\be
\widetilde{G}_{{\scriptsize{E}}}(p^2) = - \ln \Big( \frac{\lambda^2 e^2}{p^2}
\Big) - \frac{1}{2}  \ln^2 \Big( \frac{\lambda^2 e^2}{p^2} \Big) -
\frac{I_{log}^{(2)}(\lambda^2)}{b} \, ,
\ee
from which we may finally define the finite part of the two loop amplitude which
in turn satisfies a Callan-Symanzik  renormalization group equation with
renormalization scale $\lambda$.  In Minkowski space it reads:
\be
\widetilde{G}_{\mbox{{\scriptsize{fin}}}} (p^2) =     - \ln \Big(
\frac{- \lambda^2 e^2}{p^2} \Big) - \frac{1}{2}  \ln^2 \Big( \frac{-\lambda^2
e^2}{p^2} \Big) \, .
\ee

\section*{Appendix B}

The graph displayed in figure \ref{fig6} contains no
subdiagrams. The irreducible loop integral that represents the ultraviolet
divergent content of this graph can be displayed by studying $H_6^{\infty}$ in
equation   (\ref{eqn:h6}) . It will be convenient to express the
corresponding counterterm as one of the basic divergent integrals which have
appeared so far, namely $I_{log} (\lambda^2)$,   $I_{log}^{(2)} (\lambda^2)$,
$I_{log}^{(3)} (\lambda^2)$, etc.  Because this diagram is primitive and
ultraviolet logarithmically divergent,  $Z_3^6$ is expected to be
proportional do $I_{log} (\lambda^2)$. We use the same technique of appendix A
in order to extract such ultraviolet behaviour.  Moreover,  it
will become clear why the coefficient of $\zeta (3)$  in (\ref{eqn:z36}), which
will appear in the three loop  contribution to the $\beta$ function,  is
universal (save a coupling constant redefinition involving   $\zeta (3)$)
despite  the $\beta$-function being scheme dependent at the three loop level.

We temporarily work in Euclidean space where  $H_6^{\infty}$ reads
\be
H_{6 \scriptsize{E}}^{\infty} = \frac{ (4 \pi)^6}{2} \int_{l,q,k} \frac{1}{q^2
k^2  l ^2 ( l - q )^2 ( l - k )^2 (q - k)^2}  \, .
\ee
Equations (\ref{eqn:a2}) and (\ref{eqn:a3}) enable us to write
\be
H_{6 \scriptsize{E}}^{\infty} = 2 (4 \pi)^2 \int_l  \frac{1}{l^4} \,\, {\cal{Y}}
\ee
where
\bq
{\cal{Y}} &=&  \int \frac{d \Omega_q}{2 \pi^2}  \int \frac{d \Omega_k}{2 \pi^2}
\frac{1}{q \,\, k}  \sum_{m, n, p} \langle l | q \rangle^{n+1}
\langle l | k \rangle^{m+1}  \langle  q | k \rangle^{p+1} \times  \nonumber \\
&\times& C_n (l,q) \,   C_m (l,k)  \, C_p (q,k)  \, .
\label{eqn:b1}
\eq
Thus the identity
\be
\int \frac{d \Omega_k}{2 \pi^2}    C_n
\Big( \frac{l \cdot k}{l  k}\Big) C_m
\Big( \frac{q \cdot k}{q k}\Big)  = \frac{\delta_{mn}}{n+1} C_n
\Big( \frac{l \cdot q}{l  q}\Big)\, ,
\label{eqn:a2}
\ee
applied to (\ref{eqn:b1}) yields, after straightforward algebra,
\be
{\cal{Y}} = \frac{3}{2} \sum_{n=0}^{\infty}   \frac{1}{(n+1)^3} =   \frac{3}{2}
\zeta (3) \, .
\ee
Therefore, in Minkowki space,  after introducing an infrared cutoff $\mu$ and
making use of  equation   (\ref{eqn:scale}) we obtain
\be
H_{6}^{\infty} = - 3 i (4 \pi)^2 \zeta (3)  I_{log} (\lambda^2) + 3 \zeta (3)
\ln \Big( \frac{\lambda^2}{\mu^2} \Big) \, ,
\ee
where  $ \mbox{lim}_{\mu \rightarrow 0} [ 3 \zeta (3)
\ln ( \lambda^2 / \mu^2) + {\cal{F}}_6 (p^2, \mu^2)] = {\cal{G}}_6 (p^2,
\lambda^2) $ is well-defined.

\section*{Acknowledgements}

The authors wish to thank Prof. Victor Rivelles for enlightening discussions
and CNPq-Brazil  for the financial support.  M.C.N.
thanks S. Gobira for useful discussions during early stages of this work.

\end{document}